\documentclass[aps,superscriptaddress,prl,twocolumn,showpacs,10pt]{revtex4-1}
\usepackage[english]{babel}
\usepackage[utf8]{inputenc}
\usepackage{mathtools}
\usepackage{graphicx}
\usepackage{amsmath}
\usepackage{amsfonts}
\usepackage{amstext}
\usepackage{siunitx}
\newcommand{\Tr}{\mathop{\mathrm{Tr}}\nolimits}
\newcommand{\ket}[1]{\left|#1\right\rangle}
\newcommand{\bra}[1]{\left\langle#1\right|}
\newcommand{\braket}[2]{\left\langle#1\middle|#2\right\rangle}
\newcommand{\epsmean}{\varepsilon_{\text{avg}}}
\newcommand{\operator}[1]{\hat{\mathrm{#1}}}
\newcommand{\fixed}[1]{\textbf{#1}}
\begin{document}
\title{Single step quantum state engineering in traveling optical fields}
\author{Gabor Mogyorosi}
\affiliation{Institute of Physics, University of P\'ecs, Ifj\'us\'ag \'utja 6, H-7624 P\'ecs, Hungary}
\author{Peter Adam}
\affiliation{Institute of Physics, University of P\'ecs, Ifj\'us\'ag \'utja 6, H-7624 P\'ecs, Hungary}
\affiliation{Institute for Solid State Physics and Optics, Wigner Research Centre for Physics, Hungarian Academy of Sciences, P.O. Box 49, H-1525 Budapest, Hungary}
\author{Emese Molnar}
\affiliation{Institute of Physics, University of P\'ecs, Ifj\'us\'ag \'utja 6, H-7624 P\'ecs, Hungary}
\author{Matyas Mechler}
\affiliation{Institute of Physics, University of P\'ecs, Ifj\'us\'ag \'utja 6, H-7624 P\'ecs, Hungary}
\date{\today}

\begin{abstract}
We propose a general experimental quantum state engineering scheme for the high-fidelity conditional generation of a large variety of nonclassical states of traveling optical fields. It contains a single measurement, thereby achieving a high success probability. The generated state is encoded in the optimal choice of the physically controllable parameters of the arrangement. These parameter values are determined via numerical optimization. 
\end{abstract}

\maketitle

Nonclassical states of light play an essential role in numerous applications in optical quantum information processing, quantum-enhanced metrology, and fundamental tests of quantum mechanics. Measurement-induced conditional preparation is an efficient method for generating quantum states of traveling optical fields required in many of these applications. This consists in the measurement of one of the modes of a bipartite correlated state, thereby  projecting the state of the other mode to the desired one. 
Most of such conditional schemes have been developed specially for generating optical cat states. Indeed, these states and also their squeezed versions have already been prepared in several traveling wave experiments \cite{Neergaard-Nielsen2006, ourjoumtsev2007, Takahashi2008, Gerrits2010, Etesse2015, Huang2015}.

The generation of a broader class of relevant nonclassical states requires a more general approach to quantum state engineering, especially for states lacking a specialized preparation scheme. The aim of these general protocols is the preparation of arbitrary states in the same experimental setup \cite{szabo1996, DaknaPRA1999, fiurasek2005, Gerry2006, Bimbard2010, Sperling2014, adam2015, Huang2016}.

It is a common approach, for instance, to construct systematically the photon number expansion of the quantum states up to a given photon number. The methods developed for this task are based on repeated photon additions~\cite{DaknaPRA1999}, photon subtractions~\cite{fiurasek2005} and various combinations of these~\cite{Lee2010, Sperling2014}. In such schemes the number of the optical elements and detection events is generally proportional to the amount of number states involved in the photon number expansion of the target state. This property obviously leads to a decrease in the success probability and even to that in the fidelity of the preparation of states involving larger photon number components. There have been two quantum state engineering schemes proposed which can overcome this issue~\cite{MolnarPRA2018}. These contain only a few beam splitters and two or three homodyne measurements and are capable of preparing nonclassical states by the application of discrete coherent-state superpositions approximating the target states. These schemes, however, still exhibit a moderate success probability, owing to the application of multiple measurements.

In this paper we show that a single-step conditional generation scheme using separately prepared squeezed coherent states as inputs can be applied for preparing a large variety of nonclassical states with high fidelity and success probability.

\begin{figure}[!b]
	\centerline{\includegraphics[width=0.9\columnwidth]{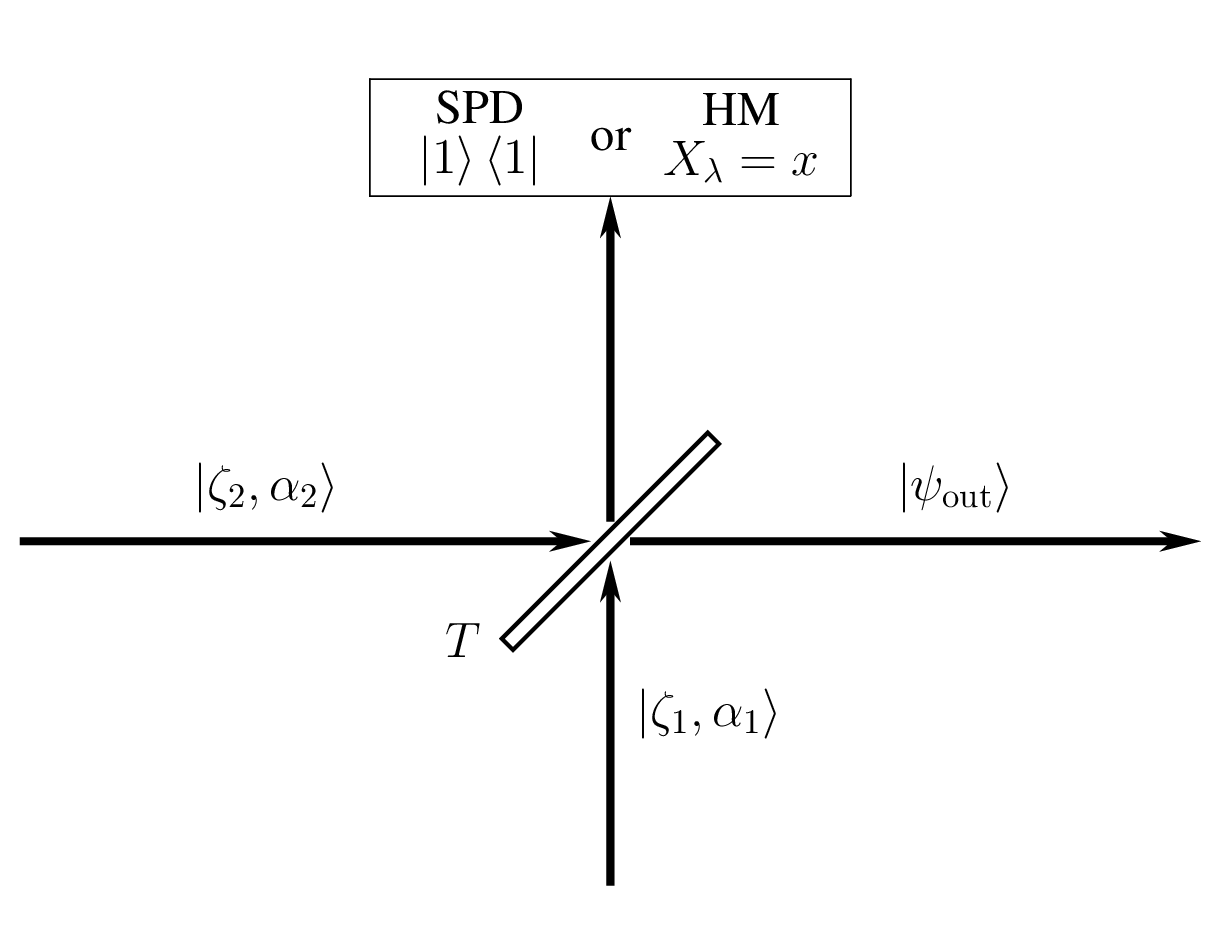}}
	\caption{Experimental scheme with two types of measurements for generating a large variety of nonclassical states in traveling optical field. The input states $\ket{\zeta_i,\alpha_i}$ interfering on a beam splitter with transmittance $T$ are squeezed coherent states.\label{fig:layout}}
\end{figure}

The proposed conditional scheme is presented in Fig.~\ref{fig:layout}. Two squeezed coherent states $\ket{\zeta_j,\alpha_j}$ with squeezing factors $\zeta_j=r_j\exp(i\theta_j)$ and coherent amplitudes $\alpha_j=|\alpha_j|\exp(i\phi_j)$ ($j=1,2$) overlap with a $\pi/2$ phase shift on a tunable beam splitter of transmittance $T$.
Then a single-photon detection (SPD) or a homodyne measurement (HM) measuring the rotated quadrature operator $X_{\lambda}$ is performed on one of the outputs to herald the generation of the desired output state $\ket{\psi_{\text{out}}}$ on the other mode. In the case of SPD the output state is non-Gaussian while for the scheme with HM it is Gaussian. 
The significance of the latter scheme is that it can vary the characteristics of the outgoing Gaussian state even at fixed inputs enabling the generation of a wide variety of states in a single setup.

The photon number expansions of the output states corresponding to the two types of measurements can be described by the expressions presented in Eqs.~\eqref{eq:mainformula2} and \eqref{eq:mainformula} for the setups containing SPD and HM, respectively, where we have introduced the function $B_p^q(x) = \binom{q}{p}\big(\sqrt{x}\big)^{q-p}\big(\sqrt{1-x}\big)^p$ and the notation $\beta_j=\alpha_j\cosh(r_j)+\alpha_j^*e^{i\theta_j}\sinh(r_j)$. The photon number expansion presented in Eq.~\eqref{eq:mainformula2} contains 9 adjustable parameters. These are: the complex squeezing parameters $\zeta_j$ and the complex coherent signals $\alpha_j$ of the input squeezed coherent states, and the transmittance $T$ of the beam splitter. The other expansion given by Eq.~\eqref{eq:mainformula} contains another two parameters additionally, namely, the measurement results $x$ of the quadrature operator $X_{\lambda}$ and the rotation angle $\lambda$ characterizing this operator.
Quantum state engineering of various target states can be realized by the proper choice of these parameters. The optimal choice can be determined by minimizing the misfit $\varepsilon = 1 - \left|\braket{\psi_{\text{out}}}{\psi_{\text{target}}}\right|^2$ between the desired state $\ket{\psi_{\text{target}}}$ and the approximated one $\ket{\psi_{\text{out}}}$ which is generated.
\begin{widetext}
	\begin{eqnarray}
    		\ket{\psi_{\text{out}}}_{\text{SPD}} &=& \mathcal{N}_{\text{out}}\prod_{j=1}^2\frac{\exp\left(-\frac{1}{2}|\alpha_j|^2-\frac{1}{2}\alpha_j^{*2}e^{i\theta_j}\tanh(r_j)\right)}{\sqrt{\cosh(r_j)}} \nonumber\\
        &&\times \sum_{n=0}^{\infty}\sum_{m=0}^{\infty}\sum_{k=0}^n i^{2k-n+1}\frac{\left(\frac{1}{2}e^{i\theta_1} \tanh(r_1)\right)^{\frac{n}{2}}}{n!}\frac{\left(\frac{1}{2}e^{\mathrm{i}\theta_2}\tanh(r_2)\right)^{\frac{m}{2}}}{m!}H_n\left(\beta_1[e^{i\theta_1}\sinh(2r_1)]^{-\frac{1}{2}}\right) \nonumber\\
		&&\times  H_m\left(\beta_2[e^{i\theta_2}\sinh(2r_2)]^{-\frac{1}{2}}\right) B_k^n(T)B_{n+m-k-1}^m(1-T)\sqrt{(n+m-1)!}~\ket{n+m-1},
		\label{eq:mainformula2}\\
		\ket{\psi_{\text{out}}}_{\text{HM}} &=& \mathcal{N}_{\text{out}} \pi^{-\frac{1}{4}} e^{-\frac{1}{2}x^2} \prod_{j=1}^2 \frac{\exp\left(-\frac{1}{2}|\alpha_j|^2-\frac{1}{2}\alpha_j^{*2}e^{i\theta_j}\tanh(r_j)\right)}{\sqrt{\cosh(r_j)}}\nonumber\\
		&&\times \sum_{n=0}^{\infty}\sum_{m=0}^{\infty}\sum_{k=0}^n\sum_{l=0}^m (-1)^l \left(\frac{1}{4}e^{i(\theta_1-2\lambda)}\tanh(r_1)\right)^{\frac{n}{2}}\left(-\frac{1}{4}e^{i(\theta_2-2\lambda)}\tanh(r_2)\right)^{\frac{m}{2}} \frac{\sqrt{(k+l)!}}{n! m!}\left(\sqrt{2}ie^{i\lambda}\right)^{k+l} \nonumber\\
		&&\times B_k^n(T)B_l^m(1-T) H_{n+m-(k+l)}(x) H_n\left(\beta_1 [e^{i\theta_1} \sinh{(2r_1})]^{-\frac{1}{2}}\right) H_m\left(\beta_2 [e^{i\theta_2} \sinh{(2r_2)}]^{-\frac{1}{2}}\right) \ket{k+l}.
	\label{eq:mainformula}
	\end{eqnarray}
\end{widetext}

In addition to misfit the probability of success is another figure of merit characterizing the performance of a conditional scheme. In the case of SPD it is defined as
\begin{math}
	P = \mathrm{Tr}(\hat{\varrho}_3\ket{1}\bra{1}),\label{eq:Prob2}
\end{math}
while for HM resulting in the measured value $x^{\text{opt}}$, it is defined as
\begin{eqnarray}
	P\left(x^{\text{opt}},\delta\right) = \int\limits_{x^{\text{opt}}-\delta}^{x^{\text{opt}}+\delta}\Tr\left(\operator{\varrho}_{3}\ket{x}{\bra{x}}\right)d x,\label{eq:Prob}
\end{eqnarray}
where $\operator{\varrho}_3 = \Tr_4\left(\ket{\psi_{\text{out}}}_{34}\prescript{}{34}{\bra{\psi_{\text{out}}}}\right)$
is the density operator of the mode on which the measurement is performed. The two-mode output state after the beam splitter is the state $\ket{\psi_{\text{out}}}_{34}$ not presented here explicitly. The parameter $\delta$ defines the range in which the misfit parameter $\varepsilon$ is assumed to be smaller than a prescribed value. As the misfit parameter changes with the measurement results within the measurement ranges, the accuracy of the preparation can be characterized by the average misfit defined as $\epsmean = \sum_{j}\varepsilon_j P_j/\sum_{j}P_j,$ where the misfits $\varepsilon_j$ and the probabilities $P_j$ are calculated for appropriately small subranges of the whole measurement range \cite{MolnarPRA2018}.

In the following we demonstrate through examples that the proposed scheme is capable of generating a wide variety of nonclassical states with high performance.
Our examples include binomial states $\ket{p,M}_{\mathrm B}$ \cite{Stoler1985}, negative binomial states $\ket{\eta,M,\varphi}_{\mathrm{NB}}$ \cite{AgarwalPRA1992}, and amplitude squeezed states $\ket{\alpha_0,u,\delta}_\text{AS}$ \cite{Adam1991}, having the following photon number expansions:
\begin{eqnarray}
	\ket{p,M}_{\mathrm B} &=& \sum_{n=0}^M\left[\binom{M}{n}p^n(1-p)^{M-n}\right]^{\frac{1}{2}}\ket{n},\label{eq:binom}\\
    \ket{\eta,M,\varphi}_{\text{NB}} &=& \mathcal{N}\sum_{n=0}^{\infty}\binom{M+n-1}{n}^{\frac{1}{2}}(\eta e^{i\varphi})^{n}\ket{n},\\
	\ket{\alpha_0,u,\delta}_\text{AS} &=& \mathcal{N} \sum_{n=0}^{\infty}\frac{\sqrt{2\pi} \alpha_0^n}{u\sqrt{n!}}\exp\left[-\frac{(\delta-n)^2}{2u^2}\right]\ket{n}.
\end{eqnarray}
For $0<p<1$ the binomial state is the superposition of the first $M+1$ photon number states.
Binomial states can be used e.g.\ for measuring the canonical phase of the quantum electromagnetic field states \cite{Pregnell2002} or they can be applied as optimal input states for communication purposes in a non-Gaussian quantum channel \cite{Memarzadeh2016}.
Negative binomial states reduce to the Susskind-Glogower phase states for $M=1$ \cite{AgarwalPRA1992,Fu1997}.
 Amplitude squeezed states contract into the coherent state $\ket{\alpha_0}$ in the limit $u\rightarrow \infty$, while in the opposite limit $u \ll 1$, an $n$-photon number state with $n=\delta$ is achieved.  Amplitude squeezed states are intelligent states of the Pegg-Barnett number-phase uncertainty relation and also of an alternative to this relation introduced as number-operator--annihilation operator uncertainty relation for a certain parameter range \cite{Adam1991, Urizar-Lanz2010, Adam2014}. Hence, they can be used for testing various uncertainty relations experimentally  \cite{Friedland2013, Yao2015}.

\begin{figure}[!tb]
\includegraphics[width=\columnwidth]{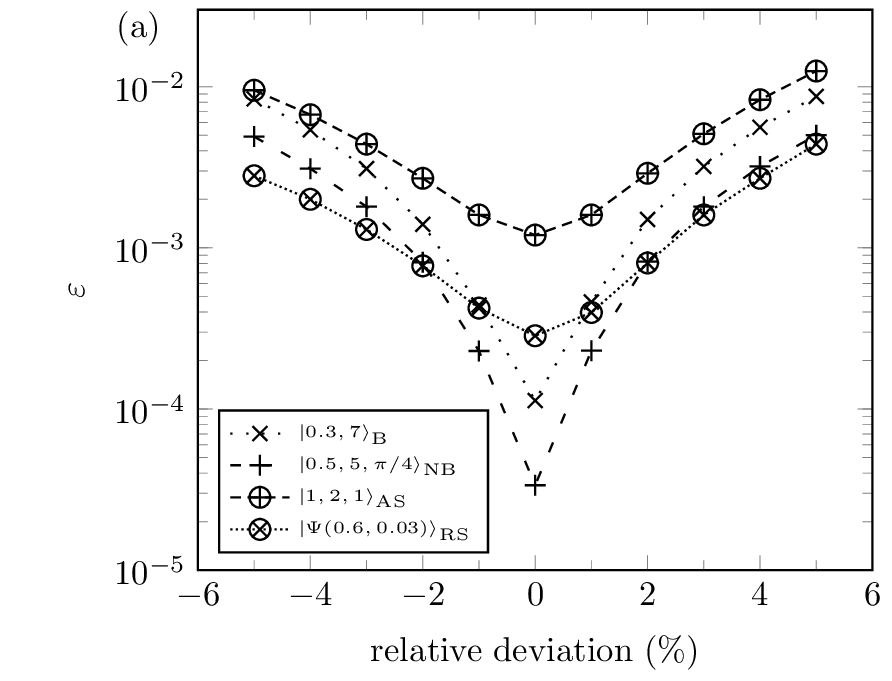}
\includegraphics[width=\columnwidth]{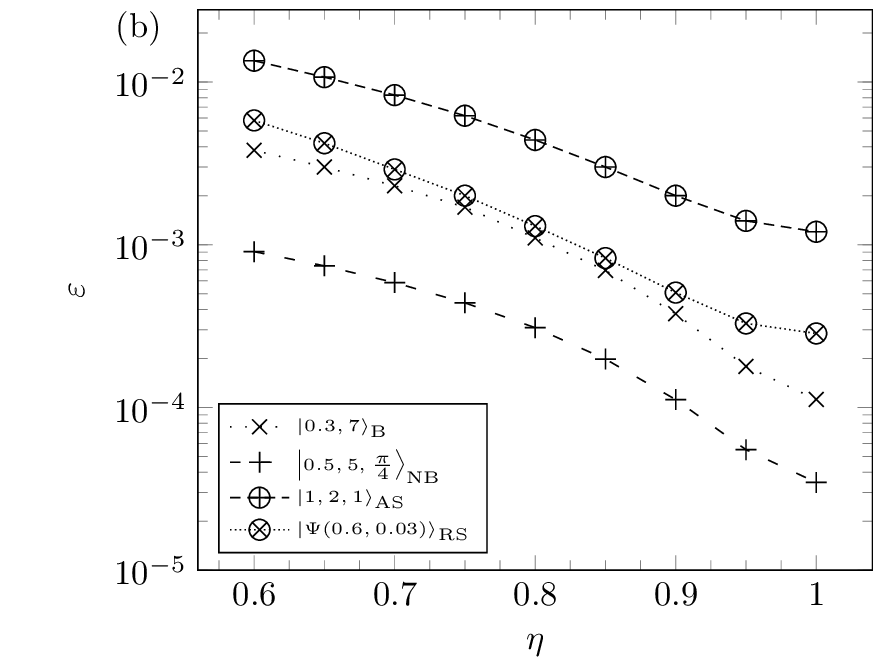}
	\caption{Misfit as a function of (a) relative deviation of the parameters of the input states from their optimal value and (b) detector efficiency $\eta$ for some target states. Binomial and negative binomial states are generated with the scheme containing HM while amplitude squeezed and resource states are generated with the scheme containing SPD.\label{fig:2}}
\end{figure}

We also consider superpositions of photon number states with \emph{ad hoc} coefficients and the special superposition
\begin{eqnarray}
	\hspace{-0.5cm}
    \ket{\psi(\zeta,\chi')}_{\text{RS}}&=&\hat{S}(\zeta)\left(\ket{0}+\chi'\frac{3}{2\sqrt{2}}\ket{1}+\chi'\frac{\sqrt{3}}{2}\ket{3}\right)\label{eq:resource}
\end{eqnarray}
referred to as resource states that can be used for realizing cubic nonlinear quantum gates essential for universal continuous-variable quantum computation in the optical setting \cite{Marek2009, Marek2011, Miyata2016}.

\begin{table*}[!tb]
	\caption{Results of the optimization for different nonclassical states for the scheme of Fig.~\ref{fig:layout}. The table presents for each state the minimal misfit $\varepsilon$ and the corresponding optimal choice of parameters: the parameters of the input squeezed coherent states ($r_1$, $\theta_1$, $\alpha_1$, $\phi_1$, $r_2$, $\theta_2$, $\alpha_2$, and $\phi_2$), the transmittance of the beam splitter $T$, and the success  probability $P$. For the HM-based version of the scheme the table also contains the measurement result $x$ of the rotated quadrature operator $X_\lambda$ rotated by the angle $\lambda$, the range $\delta$ of the measurement, and the average misfit $\epsmean$. Parameters denoted by bold characters are fixed in an \emph{ad hoc} manner in the ranges $0.3\leq r_1,r_2,\alpha_1,\alpha_2\leq 1$ and $0\leq\theta_1,\phi_1\leq\pi$ for the binomial, negative binomial, and amplitude squeezed states, and $0.4\leq r_1,r_2,\alpha_1,\alpha_2\leq 0.8$ for the resource states and the \emph{ad hoc} photon number superpositions.
    \label{tab:numerical_results}\label{tab:2}}
   \begin{center}
	\begin{tabular*}{\textwidth}{
		l@{\extracolsep{\fill}}
		S[table-format=1.2e1,table-sign-exponent]
		S[table-format=1.2]
		S[table-format=1.2]
		S[table-format=1.2]
		S[table-format=1.2]	
		S[table-format=1.2]
		S[table-format=1.2]
		S[table-format=1.2]	
		S[table-format=1.2]
		S[table-format=1.2]
		S[table-format=1.2]
		S[table-format=1.2]
		S[table-format=1.2]
		S[table-format=1.3]
		S[table-format=1.3]	}\toprule
		state & {$\varepsilon$} & {$r_1$} & {$\theta_1$} & {$\alpha_1$} & {$\phi_1$} & {$r_2$} & {$\theta_2$} & {$\alpha_2$} & {$\phi_2$} & {$T$} & {$x$} & {$\lambda$} & {$\delta$} & {$P$} & {$\epsmean$} \\\hline\\[-8pt]
		$\ket{0.3,7}_{\text{B}}$ & 1.14e-4 & \fixed{0.60} & 3.90 & \fixed{1.00} & 4.26 & \fixed{0.75} & 3.62 & \fixed{0.70} & 0.48 & 0.59 & 0.60 & 2.17 & 0.17 & 0.125 & 0.008 \\
   		$\ket{0.3,7}_{\text{B}}$ & 1.26e-4 & 0.74 & 3.50 & 0.10 & 2.14 & 0.16 & 4.43 & 1.97 & 0.08 & 0.69 &&&& 0.318 &\\
		$\ket{0.45,8}_{\text{B}}$ & 8.06e-4 & 0.45 & 0.74 & 0.34 & 1.01 & 0.45 & 0.28 & 1.97 & 0.06 & 0.90 & 0.61 & 0.04 & 0.30 & 0.275 & 0.008 \\
		$\ket{0.45,8}_{\text{B}}$ & 8.15e-4 & 0.51 & 3.22 & 2.44 & 4.95 & 0.22 & 6.18 & 0.54 & 5.58 & 0.65 &&&& 0.079 &\\
        $\ket{0.2,10}_{\text{B}}$ & 1.66e-5 & \fixed{0.60} & 1.95 & \fixed{1.00} & 4.77 & \fixed{0.75} & 2.86 & \fixed{0.70} & 6.10 & 0.49 & 0.25 & 0.56 & 0.17 & 0.132 & 0.009 \\
        $\ket{0.2,10}_{\text{B}}$ & 1.88e-5 & 0.16 & 3.39 & 0.49 & 4.70 & 0.09 & 5.68 & 1.51 & 6.27 & 0.47 &&&& 0.369 & \\
   		$\ket{0.4,15}_{\text{B}}$ & 1.91e-4 & 1.54 & 1.08 & 0.93 & 3.06 & 0.27 & 0.28 & 2.36 & 0.09 & 0.90 & 0.73 & 2.57 & 0.30 & 0.527 & 0.003 \\
		$\ket{0.65,1,0}_{\text{NB}}$ & 7.83e-4 & 0.62 & 0.13 & 0.09 & 0.25 & 0.21 & 0.90 & 0.98 & 0.02 & 0.70 & 0.23 & 0.03 & 0.20 & 0.265 & 0.008 \\
		$\ket{0.5,5,\frac{\pi}{4}}_{\text{NB}}$ & 3.36e-5 & 0.56 & 0.72 & 0.58 & 0.34 & 0.10 & 0.07 & 1.34 & 0.59 & 0.80 & 0.24 & 0.03 & 0.30 & 0.362 & 0.006 \\
		$\ket{0.5,5,\frac{\pi}{4}}_{\text{NB}}$ & 3.37e-5 & $\fixed{0.60}$ & $\fixed{1.57}$ & $\fixed{0.80}$ & $\fixed{3.14}$ & $\fixed{0.60}$ & 2.36 & 2.47 & 0.69 & 0.63 & 1.55 & 3.79 & 0.18 & 0.065 & 0.008\\
		$\ket{0.5,5,\frac{\pi}{4}}_{\text{NB}}$ & 3.40e-5 & 0.06 & 1.17 & 2.11 & 5.44 & 0.19 & 4.78 & 0.08 & 3.16 & 0.65 &&&& 0.159& \\
		$\ket{0.75,6,\frac{\pi}{2}}_{\text{NB}}$ & 3.53e-4 & \fixed{0.60} & \fixed{1.57} & \fixed{0.80} & \fixed{3.14} & \fixed{0.60} & 0.46 & 3.04 & 1.53 & 0.86 & 2.60 & 3.67 & 0.23 & 0.146 & 0.009 \\
		$\ket{0.75,6,\frac{\pi}{2}}_{\text{NB}}$ & 4.96e-4 & 0.43 & 2.45 & 0.12 & 5.57 & 0.45 & 0.32 & 3.21 & 1.63 & 0.72 &&&& 0.200 & \\
        $\ket{0.45,10,0}_{\text{NB}}$ & 8.84e-6 & \fixed{0.60} & 6.14 & \fixed{1.00} & 4.44 & \fixed{0.75} & 4.98 & \fixed{0.70} & 5.57 & 0.58 & 0.76 & 3.27 & 0.16 & 0.080 & 0.008 \\
        $\ket{0.45,10,0}_{\text{NB}}$ & 9.15e-6 & 0.08 & 5.54 & 0.07 & 2.35 & 0.12 & 3.23 & 1.69 & 0.00 & 0.88 &&&& 0.246 & \\
        $\ket{1,0.5,1}_{\text{AS}}$ & 2.45e-07 & \fixed{0.60} & 2.32 & 0.09 & 5.89 & \fixed{0.60} & 2.30 & 0.20 & 5.86 & 0.50 &&&& 0.210 &\\
        $\ket{1,0.5,1}_{\text{AS}}$ & 2.40e-07 & 0.37 & 0.68 & 0.14 & 5.02 & 0.71 & 0.67 & 0.09 & 4.98 & 0.37 &&&& 0.167&\\
		$\ket{1,1,1}_{\text{AS}}$ & 2.07e-4 & 0.45 & 1.05 & 0.76 & 5.22 & 0.50 & 0.86 & 0.42 & 5.18 & 0.51 &&&& 0.258 &\\
        $\ket{1,1,1}_{\text{AS}}$ & 2.21e-4 & \fixed{0.60} & 3.91 & 0.48 & 3.51 & \fixed{0.60} & 4.06 & 1.06 & 0.46 & 0.47 &&&& 0.270 &\\
		$\ket{1,2,1}_{\text{AS}}$ & 1.22e-3 & 0.37 & 1.61 & 1.29 & 2.40 & 0.23 & 0.86 & 1.78 & 0.36 & 0.70 & 1.71 & 3.10 & 0.40 & 0.366 & 0.007 \\
		$\ket{1,2,1}_{\text{AS}}$ & 1.18e-3 & 0.26 & 4.08 & 0.12 & 2.74 & 0.34 & 5.53 & 1.44 & 0.16 & 0.47 &&&& 0.378 &\\
		$\ket{\sqrt{3},5,3}_{\text{AS}}$ & 5.78e-5 & \fixed{0.60} & 5.14 & \fixed{1.00} & 4.53 & \fixed{0.75} & 4.61 & \fixed{0.70} & 4.72 & 0.68 & 0.79 & 2.83 & 0.16 & 0.097 & 0.008 \\
        $\ket{\sqrt{3},5,3}_{\text{AS}}$ & 1.65e-4 & 0.56 & 3.81 & 0.02 & 3.15 & 0.17 & 4.64 & 2.05 & 0.07 & 0.74 &&&& 0.389 & \\
        $\ket{1,6,1}_{\text{AS}}$ & 7.25e-7 & \fixed{0.60} & 2.49 & \fixed{1.00} & 4.22 & \fixed{0.75} & 3.09 & \fixed{0.70} & 0.47 & 0.70 & 0.87 & 4.25 & 0.17 & 0.081 & 0.009 \\
        $\ket{1,6,1}_{\text{AS}}$ & 1.49e-4 & 0.36 & 2.12 & 0.40 & 2.53 & 0.35 & 1.63 & 1.67 & 6.28 & 0.50 &&&& 0.271 &\\
		$\ket{\Psi(0.6,0.03)}_{\text{RS}}$ & 6.69e-4 & 0.46 & 2.99 & 0.07 & 6.26 & 1.15 & 0.28 & 0.02 & 1.35 & 0.30 & 0.23 & 6.13 & 0.55 & 0.222 & 0.006 \\
 		$\ket{\Psi(0.6,0.03)}_{\text{RS}}$ & 2.85e-4 & 1.02 & 2.70 & 0.76 & 5.27 & 0.61 & 0.23 & 0.36 & 4.02 & 0.79 &&&& 0.329 &\\
		$\ket{\Psi(0.15,0.1)}_{\text{RS}}$ & 7.28e-3 & 0.89 & 3.31 & 0.89 & 3.44 & 0.03 & 5.52 & 0.09 & 1.63 & 0.75 & 0.00 & 3.19 & 0.30 & 0.122 & 0.009 \\
		$\ket{\Psi(0.15,0.1)}_{\text{RS}}$ & 1.80e-3 & 1.35 & 2.78 & 0.85 & 0.3 & 0.11 & 2.81 & 0.11 & 3.77 & 0.89 &&&& 0.165 & \\
         $\ket{\Psi(0.1i,0.15)}_{\text{RS}}$ & 4.32e-3 & 0.36 & 1.64 & 0.58 & 0.60 & 0.55 & 2.30 & 0.45 & 5.23 & 0.62 &&&& 0.314 &\\
         $\ket{\Psi(0.1i,0.15)}_{\text{RS}}$ &4.74e-3 & \fixed{0.60} & 0.92 & 0.77 & 5.83 & \fixed{0.60} & 1.79 & 0.53 & 4.56 & 0.59 &&&& 0.318 &\\
 		$\ket{\Psi(0.4,0.166)}_{\text{RS}}$ & 5.31e-3 & 0.54 & 5.66 & 1.34 & 4.31 & 1.17 & 5.93 & 1.31 & 1.85 & 0.50 &&&& 0.148 &\\
        $\ket{\Psi(0.4,0.166)}_{\text{RS}}$ & 5.37e-3 &\fixed{0.60} & 1.72 & 1.14 & 5.86 & \fixed{0.60} & 1.00 & 0.96 & 4.78 & 0.54 &&&& 0.209 & \\
        $\frac{1}{\sqrt{2}}(\ket{0}+\ket{1})$ & 1.40e-6 & 0.41 & 2.52 & 0.252 & 0.63 & 0.61 & 2.52 & 0.74 & 5.88 & 0.41 &&&& 0.236 &\\
        $\frac{1}{\sqrt{2}}(\ket{0}+\ket{1})$ & 5.70e-6 & \fixed{0.60} & 0.00 & 0.82 & 4.71 & \fixed{0.60} & 6.28 & 0.25 & 3.16 & 0.50 &&&& 0.274 &\\
 		$\frac{1}{\sqrt{5}}(2\ket{1}+\ket{2})$ & 2.74e-3 & 0.35 & 6.05 & 0.41 & 4.66 & 1.39 & 6.13 & 0.21 & 0.95 & 0.35 &&&& 0.159 & \\
 		$\frac{1}{\sqrt{17}}(4\ket{1}+\ket{3})$ & 2.68e-3 & 0.71 & 5.16 & 0.01 & 1.00 & 0.79 & 4.56 & 0.00 & 0.74 & 0.46 &&&& 0.229 &\\
 		$\frac{1}{\sqrt{17}}(4\ket{1}+\ket{3})$ & 2.69e-3 & \fixed{0.60} & 1.85 & 0.00 & 4.86 & \fixed{0.60} & 2.44 & 0.00 & 2.79 & 0.60 &&&& 0.190&\\
 		$\frac{1}{3}(2\ket{0}+2\ket{1}+\ket{2})$ & 3.36e-3 & 0.19 & 5.74 & 0.76 & 4.58 & 0.27 & 6.23 & 0.22 & 0.45 & 0.72 &&&& 0.207& \\
 		$\mathcal{N}(\ket{1}+0.3\ket{3}+0.1\ket{5})$ & 7.36e-4 & 1.08 & 0.00 & 0.00 & 0.00 & 0.12 & 0.00 & 0.00 & 0.00 & 0.60 &&&& 0.131& \\
		\botrule
	\end{tabular*}
	\end{center}
\end{table*}

In order to generate a given target state, our task is to find the values of the variable parameters of the introduced scheme for which the misfit between the target and the generated states is minimal. We have used a genetic algorithm \cite{Goldberg1989} to solve this optimization problem.
We have imposed bounds on the variables so that their values are physically reasonable while the optimization problem is numerically stable and feasible. The applied ranges are $0\le r_i\le1.7$, $0\le\alpha_i\le4$, $0.1\le T\le 0.9$, $0\le x\le 4$, and all the phase angles $\theta_i$, $\phi_i$, and $\lambda$ are allowed to take any possible values between 0 and $2\pi$.

In Table~\ref{tab:numerical_results} we present the result of the optimization for several examples of the considered nonclassical states for both versions of the scheme.
In the table the range $\delta$ of the homodyne measurement has been chosen to keep the average misfit $\epsmean$ below $10^{-2}$.
We note that increasing the parameter $\delta$ increases the success probability while the fidelity decreases.

The examples show that all the considered states can be generated with higher fidelities, that is, low misfits by the proposed conditional scheme.
The achievable success probabilities of the generation are rather high compared to the ones that can be typically achieved in other quantum state engineering methods \cite{fiurasek2005, DaknaPRA1999, MolnarPRA2018}.
Binomial states and negative binomial states can be generated with similar fidelity by using either of the versions of the proposed scheme.
In the case of amplitude squeezed states the scheme yields higher fidelity with HM for low squeezing ($u\gg1$) while with SPD for high squeezing ($u\lesssim 1$), respectively.
Resource states and other special photon number superpositions can be generated by the proposed scheme with higher fidelity by using SPD than by using HM.
Generally, the states close to a Gaussian state can be generated by the scheme with HM, while the scheme containing SPD can be applied for the generation of states having a significantly non-Gaussian character.
The results also show that several nonclassical states can be approximated by Gaussian states with high fidelity.

An interesting aspect of the proposed scheme is that some of the input parameters can be chosen freely in certain ranges without the significant deterioration of the fidelity. The number, type, and ranges of such parameters can differ for different target states. The number of the fixable parameters is obviously smaller for the scheme with SPD than for the one with HM due to the smaller number of adjustable parameters.
In Table~\ref{tab:2} we show some examples of states where four or five parameters of the input states are fixed for the scheme with HM and two for the one with SPD.
This property is especially advantageous from an experimental point of view for it allows the generation of various states with high fidelity using the proposed scheme without relevant modification of the parametric setups generating the input states.

We note that most of the states presented in Table \ref{tab:numerical_results} contain higher photon-number states ($n>5$) with nonnegligible coefficients in their photon number expansion, except for the \emph{ad hoc} photon number state superpositions presented at the end of the table. Therefore these states cannot be generated realistically by quantum state engineering methods based on repeated photon addition or subtraction. The considered states can also be generated with the quantum state engineering methods based on coherent-state superpositions presented in Ref.~\cite{MolnarPRA2018}, with a significantly lower success probability though. It is important to remark that the method is suitable for generating many more states than the examples presented here. Of course in the case of certain states such as cat states for which specialized methods are also available, the general scheme may not outperform the special one.

Finally, we have considered the sensitivity of the method to the precision of the parameters of the input states and to the nonunit quantum efficiency of the measurements for both versions of the scheme.
The latter inefficiency can be represented by inserting an absorber in the signal beam path which in turn can be modeled by a fictitious beam splitter with the signal beam entering one port and the vacuum state entering the other \cite{YuenIEEE1980,LeonhardtPRA1993, BanaszekPRA1997, AppelPRA2007}. The transmittance of the beam splitter must be chosen to be equal to the quantum efficiency $\eta$. Note that such a model can describe other optical losses, e.g., the absorption of the beam path, and in the case of HM various other imperfections \cite{AppelPRA2007}. 

Fig.~\ref{fig:2} shows the change of the misfit as a function of (a) the relative shift of the parameters from their optimal value and (b) the detector efficiency $\eta$ for different target states prepared with the two versions of the scheme. One can conclude that the sensitivity is moderate for both considered inefficiencies. Hence, the considered nonclassical states can be prepared with fidelities high enough for practical applications, even using input states generated with a precision available with current experimental technology and applying realistic measurement devices.

In conclusion, we have proposed a quantum state engineering scheme based on the interference of two separately prepared squeezed coherent states for the conditional generation of a large variety of nonclassical states. Our approach unifies the benefits of simple conditional preparation and general quantum engineering schemes. It contains a single measurement thereby maintaining a proper success probability. Furthermore, it supports a broad variety of target states via parameter optimization. It can thus provide high-fidelity experimental access to many states which have relevant applications in quantum optics and quantum information science and which cannot be efficiently generated otherwise.

This research was supported by the National Research, Development and Innovation Office, Hungary (Projects No.\ K124351 and No.\ 2017-1.2.1-NKP-2017-00001 HunQuTech). The project has also been supported by the European Union, co-financed by the European Social Fund (Grants No.\ EFOP-3.6.1-16-2016-00004 entitled by Comprehensive Development for Implementing Smart Specialization Strategies at the University of P\'ecs and No.\ EFOP-3.6.2-16-2017-00005).

\bibliography{MG_Bibliography}
\end{document}